\documentclass[useAMS,usenatbib]{mn2e}

\usepackage{natbib}
\usepackage{verbatim} 
\usepackage{amsmath} 
\usepackage{amsbsy}
\usepackage{amssymb}
\usepackage{mathrsfs} 
\usepackage{pdflscape} 
\usepackage{epstopdf}
\usepackage{graphicx}
\usepackage{deluxetable}

\title[Pulsar-Black Hole Binaries in the Galactic Center]{Pulsar-Black Hole Binaries in the Galactic Center}
\author[Claude-Andr\'e Faucher-Gigu\`ere and Abraham Loeb]{Claude-Andr\'e Faucher-Gigu\`ere$^{1}$\thanks{Miller Fellow; cgiguere@berkeley.edu} and Abraham Loeb$^{2}$\thanks{aloeb@cfa.harvard.edu}\\
$^{1}$Department of Astronomy and Theoretical Astrophysics Center, University of California, Berkeley, CA 94720-3411, USA\\
$^{2}$Department of Astronomy, Harvard University, Cambridge, MA 02138, USA}

\begin{document}
\maketitle


\begin{abstract}
Binaries consisting of a pulsar and a black hole (BH) are a holy grail of astrophysics, both for their significance for stellar evolution and for their potential application as probes of strong gravity. 
In spite of extensive surveys of our Galaxy and its system of globular clusters, no pulsar-black hole (PSR-BH) binary has been found to date. 
Clues as to where such systems might exist are therefore highly desirable. 
We show that if the central parsec around Sgr A$^{\star}$ harbors a cluster of $\sim25,000$ stellar BHs (as predicted by mass segregation arguments) and if it is also rich in recycled pulsar binaries (by analogy with globular clusters), then 3-body exchange interactions should produce PSR-BHs in the Galactic center. 
Simple estimates of the formation rate and survival time of these binaries suggest that a few PSR-BHs should be present in the central parsec today. 
The proposed formation mechanism makes unique predictions for the PSR-BH properties: 1) the binary would reside within $\sim1$ pc of Sgr A$^{\star}$; 2) the pulsar would be recycled, with a period of $\sim1$ to a few tens of milliseconds, and a low magnetic field $B\lesssim10^{10}$ G; 3) the binary would have high eccentricity, $e\sim0.8$, but with a large scatter; and 4) the binary would be relatively wide, with semi-major axis $a_{\rm b}\sim0.1-\gtrsim3$ AU. 
The potential discovery of a PSR-BH binary therefore provides a strong motivation for deep, high-frequency radio searches for recycled pulsars toward the Galactic center.  
\end{abstract}


\section{Introduction}
Pulsars in binary systems provide accurate clocks that can be used to infer the properties of the binary orbit and its stellar components with precision. 
In addition to yielding important constraints on stellar evolution, binary pulsars have been used to test the validity of general relativity and to put stringent constraints on alternative theories of gravity \citep[see, e.g., the review article by][]{2004Sci...304..547S}.
The original binary pulsar PSR B1913+16 \citep[][]{1975ApJ...195L..51H}, consisting of a pulsar in orbit around another unseen neutron star, permitted the first tests of general relativity in the strong field regime and the most compelling (albeit indirect) evidence for the gravitational wave emission, via the measurement of its orbital period derivative. 
Today, the best strong-field tests of general relativity are provided by the ``double pulsar'' PSR J0737-3039 \citep[][]{2003Natur.426..531B}, the only known example of two pulsars in orbit around each other \citep[][]{2008ARA&A..46..541K}. 

While theoretical considerations suggest that a certain fraction of pulsars should have black hole (BH) companions \citep[e.g.,][]{1991ApJ...379L..17N, 1991ApJ...380L..17P, 1998A&A...332..173P, 1999ApJ...517..318B, 2002ApJ...572..962S, 2002ApJ...572..407B, 2005ApJ...628..343P} and pulsar-black hole (PSR-BH) binaries are a holy grail of pulsar astronomy, none has been found to date. 
The discovery of such a system would represent a major step forward both from the astrophysical point of view and for the unique gravity tests that it might allow. 
Clues as to where PSR-BH systems might be found are therefore highly desirable. 
Most theoretical studies have focused on PSR-BHs formed from primordial binaries in the Galactic field. 
As dense systems conducive to interactions between stars, globular clusters have also been proposed as sites where PSR-BH binaries may reside \citep[][]{2003ASPC..302..391S}. 
Indeed, the high rates of stellar encounters in globular clusters explain the overdensity of recycled millisecond pulsars\footnote{A recycled pulsar is one that has been spun up after birth by accretion of material from a binary companion.} (MSPs) and X-ray binaries in globular clusters, by orders of magnitude, relative to the field \citep[e.g.,][]{1987ApJ...312L..23V, 1991A&A...241..137H, 2000ApJ...532L..47R, 2003ApJ...591L.131P, 2006ApJ...646L.143P}. 
It is natural to expect that analogous processes might lead to the formation of PSR-BH systems. 
An important limitation of this argument, however, is that BHs may be excessively rare in globular clusters. 
To date, there is no consensus on whether globular clusters host massive central BHs similar to those commonly found at the centers of galaxies \citep[e.g.,][]{2005ApJ...634.1093G, 2008ApJ...676.1008N, 2010ApJ...710.1063V}. 
Furthermore, stellar-mass BHs are also expected to be few in globular clusters as a result of dynamical interactions that eject them \citep[][]{1993Natur.364..423S, 1993Natur.364..421K, 2004ApJ...601L.171K, 2009ApJ...690.1370M}. 

In this paper, we consider the formation of PSR-BH binaries in another kind of dense stellar environment, the center of our Milky Way. 
Indeed, the nuclear star cluster at the center of our Galaxy should allow many of the stellar interaction processes taking place in globular clusters to operate, and in addition may be particularly rich in BHs. 
In fact, the surrounding bulge provides a large reservoir of stellar remnants. 
Because stellar ($\sim10$ M$_{\odot}$) black holes are the most massive components of old stellar populations, they sink to the center owing to dynamical friction \citep[][]{1993ApJ...408..496M, 2000ApJ...545..847M, 2006ApJ...649...91F}. 
\cite{2000ApJ...545..847M} showed that about 25,000 stellar are expected to have migrated into the central parsec in this way. 
Here, we show that these BHs are likely to form binaries with pulsars via exchange interactions, and that several MSP-BH binaries should survive in the central parsec today. 
This makes the Galactic center (GC) a promising experimental target not only for pulsars orbiting the central supermassive black hole \citep[Sgr A$^{*}$;][]{1979IAUS...84..401P, 1997ApJ...475..557C, 2004ApJ...615..253P}, but also stellar-mass BHs. 
Similar processes should operate in the nuclei of other galaxies, and may actually be more efficient in low-mass galaxies with no or low-mass central black holes \citep[e.g.,][]{2009ApJ...692..917M}. 
Because of its proximity, the center of the Milky Way is however the most interesting observationally at present. 

No pulsar has been found in the central parsec so far \citep[][]{2009ApJ...702L.177D}. 
This dearth of observed pulsars in the GC is understood to arise from the strong scattering of radio waves by the dense, turbulent, ionized plasma in that region \citep[e.g.,][]{1998ApJ...505..715L, 2006MNRAS.373L...6J}, rather than by an intrinsic absence of pulsars. 
If the scattering of radio pulses broadens them by a time scale comparable to or longer than the pulse period, the signal becomes effectively constant and the pulsar disappears. 
Since the scattering broadening time scales with frequency as $\tau_{\rm scat} \propto \nu^{-4}$ \citep[e.g.,][]{1997ApJ...475..557C}, radio pulsars can nevertheless in principle be detected in the GC by observing at sufficiently high frequency. 
Recently, \cite{2010ApJ...715..939M} reported on a 15 GHz search for radio pulsars in the central parsec with the Green Bank Telescope (GBT), the highest frequency search for pulsars toward the GC to date. 
Observations at this frequency are sensitive to pulsars with periods $\gtrsim50$ ms. 
Although it was not confirmed in subsequent data taken in 2008, \cite{2010ApJ...715..939M} did detect a 607 ms pulsar candidate in their 2006 observations. 
It must be noted that the 2008 non-detection does not rule out that the candidate is a real pulsar, as a source in a short-period orbit ($\lesssim$100 yr) orbit around the GC could have its emission beam rotated away from our line of sight on a 2-yr time scale. 
Alternatively, the pulsar could have moved sufficiently in the inhomogeneous GC plasma to induce significant variations in its pulsed flux, owing either to scattering or scintillation. 
Regardless of the reality of this pulsar candidate, the upper limit of 90 ordinary pulsars in the central parsec inferred by \cite{2010ApJ...715..939M} is comparable to the expected number based on the simple expectation that there should be about as many ordinary pulsars as progenitor massive stars of mass $>8$ M$_{\odot}$ \citep[e.g.,][]{2010ApJ...708..834B}, since their lifetimes are comparable. 
Further searches are therefore highly warranted, as even current technology stands a good chance of detecting radio pulsars in the central parsec. 

For our purpose, it is especially noteworthy that even the \cite{2010ApJ...715..939M} 15 GHz periodicity search, and by extension all previous searches at lower frequencies, were completely insensitive to MSPs with pulse periods $P\approx1-$few ms. 
Existing observations thus would not have detected the MSP-BH binaries that we predict. 
Nevertheless, pulsars have been detected at a variety of higher frequencies, from 32 GHz \citep[e.g.,][]{2008A&A...480..623L}, to 43 GHz \citep[e.g.,][]{1997ApJ...488..364K}, to 87 GHz \citep[e.g.,][]{1997A&A...322L..17M}, to 144 GHz \citep[e.g.,][]{2007ApJ...669..561C}. 
Furthermore, the radio spectra of MSPs are similar to those of ordinary pulsars \citep[e.g.,][]{1998ApJ...501..270K, 1998ApJ...506..863T}, suggesting that radio observations at frequencies high enough to beat down interstellar scattering in the GC for millisecond periods are possible. 
In theory, GC MSPs could also be observed in higher-energy bands, including in the X$-$rays and $\gamma-$rays.

In the rest of this paper, we describe our new formation scenario for MSP-BH binaries in the Galactic center and present simple estimates of the formation rate and survival of these systems (\S \ref{model}). 
Because of the technical challenges of simulating the dynamics of the GC, including the critical binary processes and the effects of different stellar masses \citep[e.g.,][]{2006ApJ...649...91F, 2009ApJ...700.1933H}, and since substantial uncertainties exist, we focus here on basic analytic considerations. 
Our here is to outline the relevant physical processes and to combine them in order-of-magnitude estimates for the number of MSP-BH binaries that should survive in the GC today, as a motivation for observational efforts as well as further theoretical studies. 
In \S \ref{discussion}, we discuss the prospects for detecting the predicted systems, the unique signatures of the proposed formation channel, and the potential implications for physics and astrophysics. 

\section{MSP-BH Binaries in the GC}
\label{model}
The MSP-BH binary formation scenario that we envision in the GC is as follows. 
We assume that, in analogy to the phenomenon in globular clusters, the high stellar densities in the central parsec lead to the dynamical formation of MSPs. 
We further assume, again by analogy with globular clusters, that most of these MSPs will be found in binaries with white dwarf (WD) companions. 
In \S \ref{MSP WD progenitors}, we discuss how the properties of globular clusters are scaled to the nuclear star cluster. 
What is unique about the GC is that the MSP-WD binaries have a significant probability of undergoing a 3-body interaction with a stellar BH in the central cluster. 
A likely outcome of many such interactions is the exchange of the WD companion for the stellar BH, yielding a MSP-BH binary. 
In what follows, we investigate this scenario by quantifying the time scales for the formation and survival of the MSP-BHs.  

To simplify the calculations, we assume a steady-state stellar background and calculate the rates of MSP-BH formation and destruction in this background. 
This approximation is justified by the typical lifetimes less than a few Gyr of the MSP-BH binaries (\S \ref{survival time} and \S \ref{sensitivity to model parameters}), i.e. significantly less than the age of the Galaxy and therefore presumably of the nuclear cluster. 
Potential effects arising from recent events are outlined in \S \ref{caveats}. 

\subsection{Stellar Distribution in the GC}
\label{stellar distribution}
The supermassive black hole at the center of our Galaxy, Sgr A$^{*}$, is taken to be at the origin and to have a mass $M_{\rm SMBH}=4\times10^{6}$ M$_{\odot}$ \citep[e.g.,][]{2008ApJ...689.1044G, 2009ApJ...692.1075G}. 
\cite{2009A&A...502...91S} examined the proper motions of stars out to a distance of 1 pc from Sgr A$^{*}$ and estimated the extended mass (excluding that of the SMBH) to be $0.5-1.5\times10^{6}$ M$_{\odot}$ within that radius. 
We assume that this extended mass follows a power-law cusp
\begin{equation}
\label{stellar density profile}
\rho_{\star}(r) = \rho_{\star,0} \left( \frac{r}{r_{0}} \right)^{-\gamma},
\end{equation}
normalized such that $M_{\star}(<1{\rm pc}) = 4\pi \int_{0}^{\rm 1pc} dr r^{2} \rho_{\star}(r)=10^{6}$ M$_{\odot}$. 
The power-law index $\gamma$ is not uniquely constrained observationally \citep[][]{2009A&A...502...91S} but consistent with the value $\gamma=1.3$ predicted by \cite{2006ApJ...649...91F} for main sequence stars in their standard Milky Way nucleus simulation. 
We adopt this value in our calculations, but note that it has a small impact on our results since the SMBH dominates the velocity dispersion throughout most the region of interest ($r \lesssim 0.5$ pc). 
The gravitational potential of the combination of the SMBH and the extended central mass distribution is then
\begin{equation}
\label{SMBH power law potential}
\Phi(r) - \Phi(r_{0}) = \frac{v_{c,\rm ext}^{2}(r_{0}) - v_{c,\rm ext}^{2}(r)}{\gamma - 2} - \frac{G M_{\rm SMBH}}{r}~~~(\gamma \neq 2),
\end{equation}
where 
\begin{equation}
v_{c,\rm ext}^{2}(r) = \frac{4 \pi G \rho_{\star,0} r_{0}^{\gamma}}{3-\gamma} r^{2-\gamma}.  
\end{equation}
For a spherical system with an isotropic velocity distribution at each point, the Jeans equation reads
\begin{equation}
\frac{d(n \sigma^{2})}{dr} = -n \frac{d\Phi}{dr},
\end{equation}
where $n$ is the number density of the constituents of interest and $\sigma$ is their velocity dispersion \citep[e.g.,][]{2008gady.book.....B}. 
For the potential in equation (\ref{SMBH power law potential}) and a stellar population following $n\propto r^{-\alpha}$, this yields
\begin{equation}
\label{velocity dispersion}
\sigma_{\star}^{2}(r) = \frac{4 \pi G \rho_{\star,0} r_{0}^{\gamma}}{(3 - \gamma)(\gamma+\alpha-2)} r^{2-\gamma} + \frac{G M_{\rm SMBH}}{(1 + \alpha)r}. 
\end{equation}

If gravitational encounters between stars are sufficiently frequent, they efficiently exchange energy. 
In equipartition, this would imply that more massive stars should have a lower velocity dispersion than less massive ones by a factor $\sqrt{M_{1}/M_{2}}$, where the $M_{i}$ are the masses of two stellar species under consideration. 
Self-gravitating systems, however, are subject to the counter-acting effect of mass segregation, which tends to bring the more massive components inward, thereby leaving them on higher-velocity orbits \citep[e.g.,][]{2007MNRAS.374..703K}. 
Within the radius of influence of the SMBH, the actual squared velocity dispersion of a given species scales instead with the power-law index of its number density distribution (the $[1+\alpha]^{-1}$ factor in the second term of eq. (\ref{velocity dispersion})). 
The dispersion of the \emph{relative} velocities between any two species, relevant for interaction cross sections, is then simply $\sigma_{\star,\rm rel}=\sqrt{\sigma_{\star,1}^{2}+\sigma_{\star,2}^{2}}$. 

Because the relaxation time is short at the Galactic center, old stars are expected to have settled in a cusp of approximately power-law form \citep[e.g.,][]{1972ApJ...178..371P, 1976ApJ...209..214B, 1977ApJ...216..883B}. 
We therefore use simple power laws to describe the spatial distributions of stars. 
Of particular importance for our problem, stellar BHs of mass $\sim10$ M$_{\odot}$ from the surrounding bulge should sink into the central parsec on a dynamical friction time scale, which is about 10 Gyr (roughly the age of the Galaxy) at a radius of 5 pc from Sgr A$^{\star}$\citep[][]{1993ApJ...408..496M, 2000ApJ...545..847M}.\footnote{As the stellar BHs migrate inward, they eventually dominate the stellar distribution and their further evolution is determined by interactions with each other, rather than dynamical friction \citep[e.g.,][]{2006ApJ...649...91F}.}
This process acts to concentrate $\sim25,000$ stellar BHs in the central parsec. 
Since it takes $\approx30$ Gyr for these BHs to be swallowed by Sgr A$^{\star}$ \citep[][]{2000ApJ...545..847M}, most of these stellar BHs should remain there today.  
We parameterize the number density distribution of stellar BHs by
\begin{equation}
n_{\rm BH}(r) = n_{\rm BH,0} \left( \frac{r}{r_{0}} \right)^{-\alpha_{\rm BH}}, 
\end{equation}
where we conventionally define $r_{0}=0.5$ pc (characteristic of the size of the stellar BH cluster) and the constant $n_{\rm BH,0}$ is set by normalizing to the total number of BHs, $N_{\rm BH}$. 
The other stellar population of greatest interest for us are MSPs, whose distribution we parameterize as 
\begin{equation}
n_{\rm MSP}(r) = n_{\rm MSP,0} \left( \frac{r}{r_{0}} \right)^{-\alpha_{\rm MSP}}.
\end{equation}

For a single-mass stellar population, \cite{1976ApJ...209..214B} showed that the cusp approaches a universal slope of $-7/4$. 
In a more realistic stellar population consisting of multiple masses, \cite{1977ApJ...216..883B} argued that the slope would depend weakly on the mass of the component of interest, $\alpha = 3/2 + p$, where $p \approx 0.3 M_{i}/M_{1}$ and $M_{1}$ is the mass of the heaviest stellar component. 
In our fiducial calculations, we assume that the MSP-WD spatial distribution follows that of neutron stars, and set $\alpha_{\rm BH}=1.75$ and $\alpha_{\rm MSP}=1.3$, also consistent with the standard Milky Way nucleus simulation of \cite{2006ApJ...649...91F}. 

Two variations about these fiducial choices warrant special consideration. 
First and most importantly, some observations suggest that the stellar distribution near Sgr A$^{\star}$ may not be cuspy but instead form a core or even a cavity within $\sim0.2$ pc \citep[][]{2009ApJ...703.1323D, 2009A&A...499..483B}. 
These observations however currently only directly probe late-type red giants, and so do not necessarily exclude the presence of a relaxed cusp of stellar remnants as predicted theoretically.\footnote{For instance, physical collisions could preferentially deplete giants from the inner cusp \citep[e.g.,][]{2009MNRAS.393.1016D}.} 
Nevertheless, we will explore this possibility by truncating the PSR-BH formation rate at minimum radii $R_{\rm min}=0.1$ and 0.2 pc (\S \ref{sensitivity to model parameters}), and show that our results are only weakly dependent on the presence of an inner cusp. 
Second, there is the ``strong'' segregation regime discussed by \cite{2009ApJ...697.1861A}, in which the most massive components contribute only a minor fraction of the gravitational potential and concentrate more strongly owing to dynamical friction off less massive components in the stellar cusp \citep[see also][]{2009ApJ...698L..64K, 2010ApJ...708L..42P, 2010GWN.....3....3P, 2010arXiv1010.5781A}. 
It is possible that the Milky Way nucleus marginally satisfies the criteria for strong segregation \citep[][]{2006ApJ...645L.133H}, in which case a value as steep as $\alpha_{\rm BH}\approx2$ could be realized. 
Our simple arguments are unfortunately not well suited to treat this singular case, in which the number of BHs inside a radius $r$ is divergent for $\alpha_{\rm BH}\geq2$ if we do not account for an inner turn over. 
Furthermore, recent results suggest that resonant relaxation could carve out a cavity similar to that observed in late-type stars in the inner $\sim0.2$ pc \citep[][]{2010arXiv1010.1535M}, so that the presence of such a singular inner cusp is not at present well supported for the Milky Way. 
We will therefore not consider this possibility further in this work. 

\begin{figure*}
\mbox{
\includegraphics[width=0.5\textwidth]{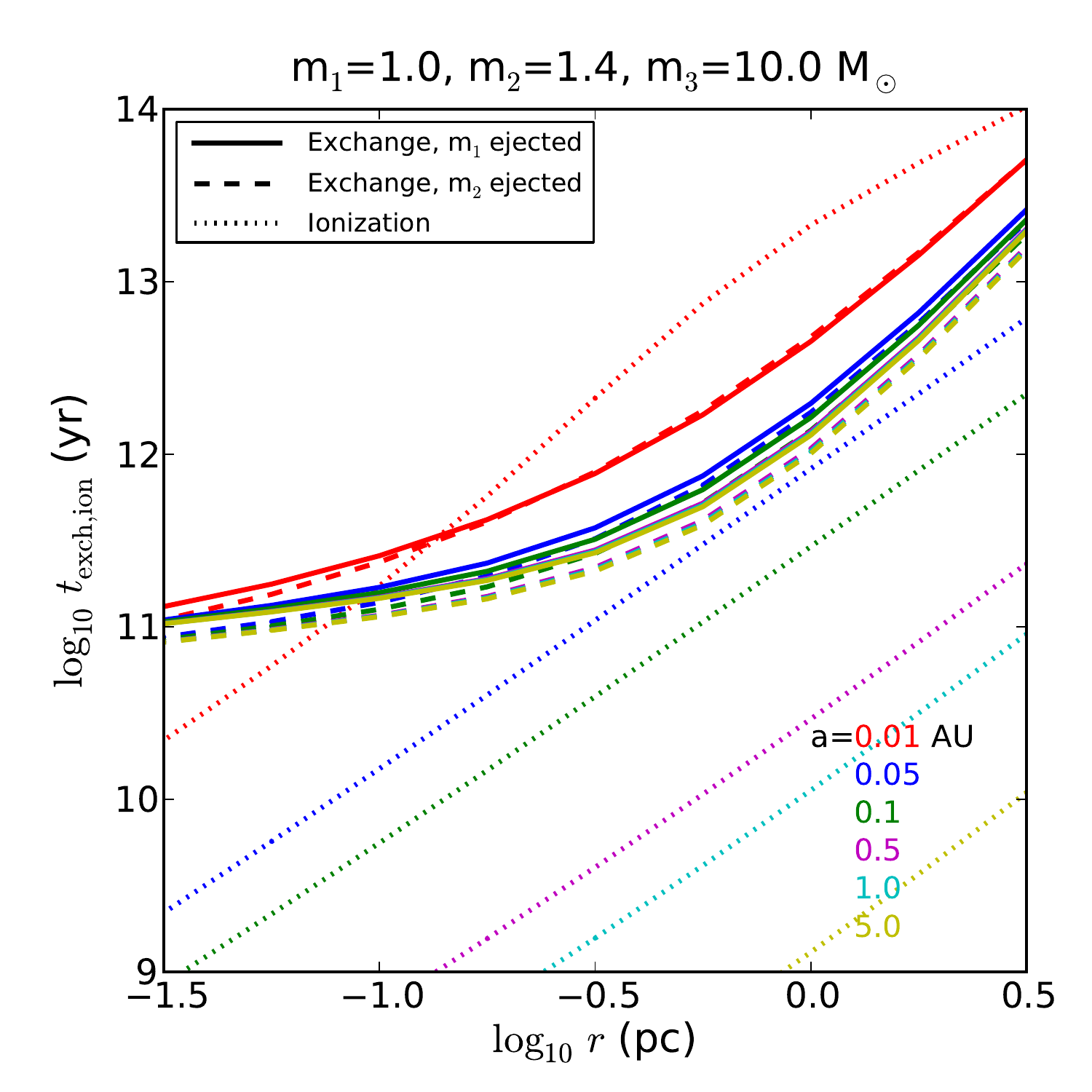}
\includegraphics[width=0.5\textwidth]{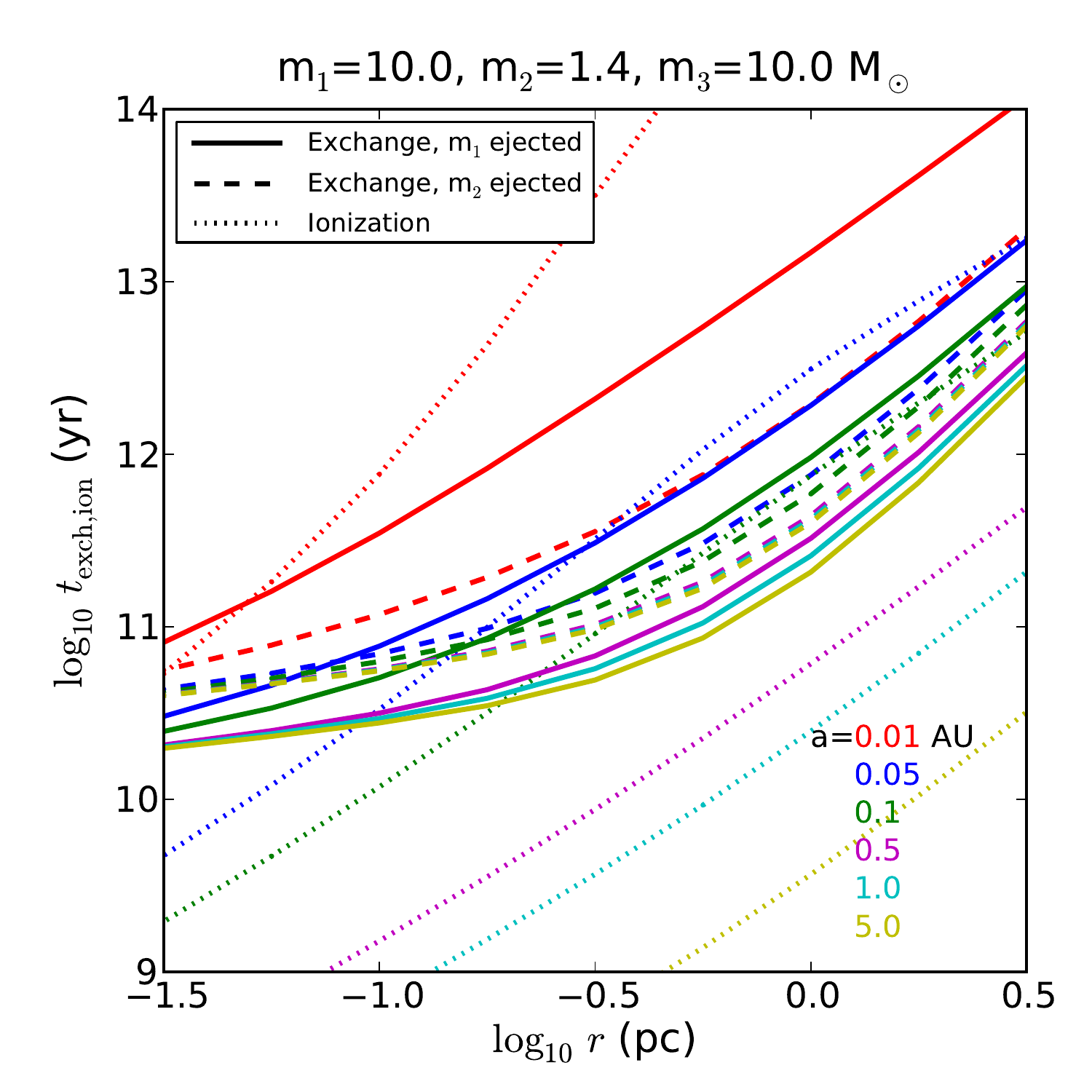}
}
\caption{Time scales for 3-body interactions of (single) binaries with a nuclear cluster of $M_{\rm BH}=m_{3}=10$ M$_{\odot}$ BHs, as a function of radius from Sgr A$^{\star}$ and for different binary semi-major axes. \emph{Left:} MSP-WD binaries with $M_{\rm WD}=m_{1}=1$ M$_{\odot}$ and $M_{\rm NS}=m_{2}=1.4$ M$_{\odot}$. \emph{Right:} MSP-BH binaries with $M_{\rm BH}=m_{1}=10$ M$_{\odot}$ and $M_{\rm NS}=m_{2}=1.4$ M$_{\odot}$. 
These plots assume the stellar distribution model for the Galactic center described in \S \ref{stellar distribution}, i.e. a central SMBH of mass $M_{\rm SMBH}=4\times10^{6}$ M$_{\odot}$ surrounded by a cluster with total mass $M_{\star}(<1~{\rm pc})=10^{6}$ M$_{\odot}$ within 1 pc and a power-law density slope $\gamma=1.3$. 
The stellar BHs are assumed to have a number density $n_{\rm BH,0}=10^{4}$ pc$^{-3}$ at $r_{0}=0.5$ pc, and radial profile of slope $\alpha_{\rm BH}=1.75$. For each process (line style), the time scale increases with decreasing semi-major axis.
}
\label{3body time scales}
\end{figure*} 

\subsection{MSP-BH Formation Rate}
\label{formation scenario}
Most of the theoretical work on 3-body encounters has focused either on the very soft regime \citep[in which the binding energy of the binaries is much smaller than the kinetic energy of background stars; e.g.,][]{1983ApJ...268..342H}, the very hard regime \citep[in which the binding energy of the binaries greatly exceeds the kinetic energy of the other stars; e.g.][]{1996ApJ...467..359H}, or on situations in which all the stars of have identical mass \citep[e.g.,][]{1983ApJ...268..319H, 1984ApJS...55..301H}. 
For our problem, many binaries straddle the soft-hard transition and the principal stellar components (WDs, PSRs, and BHs) have substantially multiple masses, which has an important impact on the reaction rates \citep[e.g.,][]{1993ApJ...415..631S}. 
To ensure accuracy, we rely on numerically evaluated 3-body cross sections that are valid regardless of the hardness of the binaries, and which incorporate the effects of different masses. 
To do so, we use the \verb+sigma3+ program from the publicly-available Starlab package \citep[][]{1996ApJ...467..348M}. 
This program automatically performs a series of direct 3-body numerical experiments and determines the cross sections using a Monte Carlo technique. 
We then evaluate the interaction rates $\langle \sigma v \rangle$ (where $\sigma$ denotes the cross section and $v$ denotes the relative velocity between the binary center of mass and the third star) by averaging over a Maxwellian velocity distribution at each radius from Sgr A$^{\star}$. 
The accuracy parameters of the automated \verb+sigma3+ program and of the integration over the Maxwellian distribution are increased until convergence is attained, with negligible resultant statistical errors. 
For simplicity, we assume that all the WDs have a fixed mass $M_{\rm WD}$, that all the neutron stars have a mass $M_{\rm NS}$, and that all the stellar BHs have a mass $M_{\rm BH}$. 

For reference, Figure \ref{3body time scales} shows 3-body interaction time scales, 
\begin{equation}
\label{time scales def}
t_{\rm exch,ion}(r) \equiv \frac{1}{n_{\rm BH}(r)\langle \sigma_{\rm exch,ion} v \rangle (r)},
\end{equation}
for relevant combinations of stellar component masses. In both panels, the interactions are assumed to take place between binaries and external stellar BHs of mass $M_{\rm BH}=10$ M$_{\odot}$, with a power-law density profile of index $\alpha_{\rm BH}=1.75$ and $n_{\rm BH,0}=10^{4}$ pc$^{-3}$ at $r_{0}=0.5$ pc, and for the nuclear mass distribution model described in the previous section. 
If the BH cluster were more concentrated in reality, the time scales would be correspondingly shorter, and vice versa. 

We first consider the spatially-averaged rate at which an individual MSP-WD system undergoes an exchange interaction yielding a MSP-BH system, for a fixed semi-major axis of the original binary, $a_{\rm b}$ (the ``creation rate''): 
\begin{align}
\label{MSP BH creation rate}
C_{\rm MSP-BH}(& M_{\rm WD},~M_{\rm NS},~M_{\rm BH},~a_{\rm b}) = \notag \\
& N_{\rm MSP}^{-1}(R_{\rm min}<r<R_{\rm max})\notag \\ 
& \times 4 \pi \int_{R_{\rm min}}^{R_{\rm max}} dr r^{2} n_{\rm MSP}(r) n_{\rm BH}(r) 
\notag \\
& \times
\langle
\sigma_{\rm exch}^{\rm WD} v
\rangle(M_{\rm WD},~M_{\rm NS},~M_{\rm BH},~a_{\rm b},~r),
\end{align}
where $N_{\rm MSP}(R_{\rm min}<r<R_{\rm max}) = 4\pi \int_{R_{\rm min}}^{R_{\rm max}} dr r^2 n_{\rm MSP}(r)$ is the total number of MSP-WD systems, and $\sigma_{\rm exch}^{\rm WD}$ is the cross section for an exchange interaction that ejects the WD and leaves behind a MSP-BH binary. 
We implicitly assume that all MSPs have a WD companion. 
$R_{\rm min}$ and $R_{\rm max}$ correspond to the minimum and maximum radii at which MSP-BH formation proceeds. 
Very close to the SMBH, binaries are tidally separated, so that $R_{\rm min}>0$; we will however show that our results are weakly dependent on the exact value of $R_{\rm min}$ and fiducially adopt $R_{\rm min}=0.1$ pc. 
$R_{\rm max}$ simply corresponds to the outer radius of the stellar BH cluster in which the exchange interactions take place. 
In their simple analytic modeling, \cite{2000ApJ...545..847M} predict $R_{\rm max}=0.7$ pc, while the numerical calculations of \cite{2006ApJ...649...91F} suggest that the BH density profile has already steepened substantially at $r\approx0.5$ pc. 
We adopt this latter value for $R_{\rm max}$ in our fiducial estimates. 

As Figure \ref{3body time scales} shows, the 3-body interaction cross sections are sensitive to the binary semi-major axis.
Because semi-major axes typically have wide distributions, it is important to average the creation rate over such a distribution. 
We assume that the semi-major axis distribution is flat in logarithmic space between specified limits: 
\begin{equation}
\frac{dP}{d\ln{a_{\rm b}}} = \left\{ 
\begin{array}{cl}
\frac{1}{\ln{a_{\rm b,max}}-\ln{a_{\rm b,min}}} & a_{\rm b,min}<a_{\rm b}<a_{\rm b,max} \\
0 & {\rm otherwise} \\
\end{array}
\right. .
\end{equation}
This simple analytic form is a crude approximation to the observed distribution of binary pulsar semi-major axes and is a reasonable first-order approximation for our exploratory estimates at the GC, where there are currently no empirical constraints. 
We adopt $(a_{\rm b,min},~a_{\rm b,max})=(0.01,~0.3)$ AU in our fiducial calculations. 
The lower limit corresponds roughly to the orbital separation at which circular MSP-WD binaries merge in $<1$ Gyr owing to the emission of gravitational waves (see eq. (\ref{gw time scale}) below).
The upper limit is roughly the maximum separation of MSP-WD binaries that survive evaporation for $>1$ Gyr (Fig. \ref{3body time scales} and \S \ref{survival time}).  

We define:
\begin{equation}
\langle C_{\rm MSP-BH}\rangle_{a_{\rm b}} \equiv \int_{a_{\rm b,min}}^{a_{\rm b,max}} d\ln{a_{\rm b}} \frac{dP}{d\ln{a_{\rm b}}} C_{\rm MSP-BH}(a_{\rm b}).
\end{equation}
In the absence of destruction processes and assuming a steady creation rate, the total number of MSP-BH systems formed after a time equal to the age of the Galaxy, $t_{\rm Gal}$, would be:
\begin{equation}
N_{\rm MSP-BH}^{\rm upper} = \langle C_{\rm MSP-BH}\rangle_{a_{\rm b}} t_{\rm Gal} N_{\rm MSP}.
\end{equation}
For our fiducial parameter values and $N_{\rm BH}=25,000$ stellar BHs between $R_{\rm min}=0.1$ pc and $R_{\rm max}=0.5$ pc,
\begin{equation}
\frac{N_{\rm MSP-BH}^{\rm upper}}{N_{\rm MSP}} 
\approx 0.04 \left( \frac{ \langle C_{\rm MSP-BH}\rangle_{a_{\rm b}}}{ \langle C_{\rm MSP-BH}\rangle_{a_{\rm b}}^{\rm fid}} \right) \left( \frac{t_{\rm Gal}}{10~{\rm Gyr}} \right),
\end{equation}
suggesting that up to a few percent of MSPs in the GC could be in BH binaries.

\subsection{Survival Time of the MSP-BH Binaries}
\label{survival time}
In practice, the MSP-BH binaries are continuously destroyed. 
A more realistic estimate of the number remaining at the present time is therefore
\begin{equation}
\label{N surv}
N_{\rm MSP-BH}^{\rm surv} = \langle C_{\rm MSP-BH} t_{\rm surv} \rangle_{\rm a_{\rm b}} N_{\rm MSP},
\end{equation}
where $t_{\rm surv}$ is the time for which a given binary survives. 
Equation (\ref{N surv}) is easily understood as the number of MSP-BHs created less than a time $t_{\rm surv}$ in the past, and the survival time can be expressed as the minimum time of the destruction time scales for the different relevant processes. 

We now summarize the different processes that can destroy the MSP-BH binaries in the GC and their corresponding time scales.\\ \\
\emph{Ionization:} Encounters with field stars can ionize the MSP-BH binaries. 
Within the cluster of stellar BHs, binary ionization is dominated by encounters with the BHs owing to the $m_{3}^{2}$ dependence of the ionization cross section, where $m_{3}$ is the mass of the third star \citep[e.g.,][]{1983ApJ...268..342H}. 
This is because the ionization rate scales as $\rho_{3} m_{3}$ (where $\rho_{3}=n_{3}m_{3}$ and $n_{3}$ is the number density), and this factor is largest by a factor of several for BHs in the standard Milky Way model of \cite{2006ApJ...649...91F} on which we base our calculations. 

Ionization proceeds in two regimes. 
First, the binaries can be ionized in single encounters. 
In the limit of a high-velocity encounter, the time scale for this process is
\begin{equation}
\label{ionization time scale}
t_{\rm ion} = \frac{3}{80}
\left( \frac{2}{\pi} \right)^{1/2}
\frac{\sigma_{3}}{G a_{\rm b'} \rho_{3}}
\frac{m_{1}+m_{2}}{m_{3}},
\end{equation}
where we have defined $t_{\rm ion}=(n_{3} \langle \sigma_{\rm ion} v \rangle)^{-1}$ and $\sigma_{3}$ is the relative velocity dispersion between the third bodies and the centers of mass of the $m_{1},~m_{2}$ binaries. 
The cross section $\sigma_{\rm ion}$ for ionization was derived by \cite{1983ApJ...268..342H}, and we averaged over a Maxwellian velocity distribution. 
We use a primed $a_{\rm b}'$ to denote the semi-major axis of the MSP-BH formed in the exchange interaction, which in general differs from the semi-major axis of the original binary. 

The cumulative effect of weak encounters can also gradually evaporate the binaries. 
Generalizing the Fokker-Planck derivation in \cite{2008gady.book.....B} for the case of equal masses, the time scale for the evaporation process is
\begin{align}
\label{evaporation time scale}
t_{\rm evap} = \frac{1}{16}
\left( \frac{3}{\pi} \right)^{1/2}
& \frac{\sigma_{3}}{G a_{\rm b}' \rho_{3} \ln{\Lambda}}
 \frac{m_{1} m_{2}}{m_{3}^{2}} \\
& \times \frac{(m_{1} + m_{3})(m_{2} + m_{3})}{2m_{1}m_{2} + m_{3}(m_{1} + m_{2})},
\end{align}
where
\begin{equation}
\Lambda \approx
\left( \frac{8}{\pi} \right)^{1/2} 
\frac{a_{\rm b}' (\sigma_{12}^{2} + \sigma_{3}^{2})}{2G[\min{(m_{1},~m_{2})} + m_{3}]}.
\end{equation}
For $m_{1}=M_{\rm BH}$, $m_{2}=M_{\rm NS}$, and $m_{3}=M_{\rm BH}$, appropriate for MSP-BHs in a cluster of stellar BHs, $t_{\rm evap}\approx0.4 t_{\rm ion} / \ln{\Lambda}$. 
For characteristic values $a_{\rm b}'=1$ AU and $\sigma_{12}=\sigma_{3}=200$ km s$^{-1}$, $\Lambda\approx7.2$ and so $t_{\rm evap}\approx0.2t_{\rm ion}$, only logarithmically dependent on the binary semi-major axis and surrounding velocity dispersion. 
The diffusive evaporation process therefore dominates over discrete ionization, and so we use a tidal separation time scale of $t_{\rm evap}=0.2t_{\rm ion}$ in our calculations. 
This is a conservative approximation, as $t_{\rm evap}$ is actually within a factor $\sim2$ of $t_{\rm ion}$ for the binaries that are only marginally soft \citep[][]{1983ApJ...272L..29H, 1995ApJ...438L..33R}.
\\ \\
\emph{Exchange:} When MSP-BHs encounter other stars, an exchange interaction may take place that ejects one of the binary components. 
As for ionization, this process is dominated by encounters with other stellar BHs, and in this case only the exchanges that eject the MSP effectively destroy the MSP-BH. 
The time scale for this process is evaluated as in equation (\ref{time scales def}), using cross sections numerically computed with Starlab. 
In most cases, exchanges are negligible relative to binary evaporation (see the right panel of Fig. \ref{3body time scales}).\\ \\
\emph{Gravitational wave-driven merger:} A MSP-BH binary will merge owing to the emission of gravitational radiation after a time
\begin{align}
\label{gw time scale}
t_{\rm GW} \approx 200~{\rm Gyr}~
&\left(
\frac{a_{\rm b}'}{0.1~{\rm AU}}
\right)^{4}
\left(
\frac{M_{\rm NS}+M_{\rm BH}}{11.4~{\rm M_{\odot}}}
\right)^{-1} \notag \\
&\times \left(
\frac{M_{\rm NS}M_{\rm BH}}{14~{\rm M_{\odot}^{2}}}
\right)^{-1}
 (1 - e'^{2})^{7/2},
\end{align}
where $e'$ is its eccentricity of the MSP-BH formed in the exchange interaction \citep[][]{1964PhRv..136.1224P, 2005ApJ...628..343P}. 
In the hard limit (where the exchange cross section is important), the semi-major axis of the new binary can be estimated as $a_{\rm b}'\sim a_{\rm b} (M_{\rm BH}/M_{\rm WD})$ \citep[][]{1993ApJ...415..631S}
and the new binary eccentricity is high, since $1-e'\sim R_{\rm p}/a_{\rm b}'$, where $R_{\rm p}\sim a_{\rm b}'(M_{\rm WD}/M_{\rm BH})[(M_{\rm WD} + M_{\rm NS} + M_{\rm BH})/(M_{\rm WD} + M_{\rm NS})]^{1/3}$ \citep[][]{1996ApJ...467..359H}. 
We use these analytic approximations for the semi-major axis and eccentricity in our calculations. 
In the limit $M_{\rm WD} \ll M_{\rm NS} \ll M_{\rm BH}$, the eccentricity expression simplifies to $1-e'\sim(M_{\rm WD}/M_{\rm BH})(M_{\rm BH}/M_{\rm NS})^{1/3}$, indicating that the key parameter is $M_{\rm WD}/M_{\rm BH}$. 
For our fiducial masses, $a_{\rm b}'\sim10a_{\rm b}$ and $e'\sim0.8$. 
Because of the strong scaling of $t_{\rm GW}$ with $a_{\rm b}'$, and since the semi-major axis is increased by a factor $M_{\rm BH}/M_{\rm WD}\sim10$ in the exchange interaction, gravitational wave-driven merging is ultimately unimportant.\\ \\
\emph{Tidal separation:} If a binary with component masses $m_{1}$, $m_{2}$, and semi-major axis $a_{\rm b}$ gets within a distance 
\begin{align}
\label{tidal radius}
r_{T} = a _{\rm b} & \left( \frac{M_{\rm SMBH}}{m_{1}+m_{2}} \right)^{1/3} \approx 3.3\times10^{-5}~{\rm pc}
\left( \frac{a_{\rm b}}{0.1~{\rm AU}} \right) \notag \\
& \times \left( \frac{m_{1}+m_{2}}{11.4{\rm~M_{\odot}}} \right)^{-1/3}
\left( \frac{M_{\rm SMBH}}{4\times10^{6}~{\rm M_{\odot}}} \right)^{1/3}
\end{align}
of the central massive black hole, the tidal force will separate it. 
Tidal separation in the central parsec occurs predominantly via the loss cone, i.e. through angular momentum diffusion which can shrink the periapsis distance from the SMBH by increasing the orbital eccentricity \citep[e.g.,][]{1976MNRAS.176..633F, 1977ApJ...211..244L, 2003ApJ...599.1129Y}. 

A simple estimate of the time scale for separating MSP-BH binaries via the loss cone is obtained by noting that the process is analogous to the one responsible for the direct swallowing of stellar BHs by the SMBH, but with a larger destruction inner radius $q_{\rm min}$ equal to $r_{T}$. 
\cite{2000ApJ...545..847M} showed that the time scale for depleting the stellar BH cluster by direct swallowing is $t_{\rm lc}\approx30$ Gyr and scales only logarithmically with $q_{\rm min}$ as
\begin{equation}
\label{loss cone log factor}
t_{\rm lc}
\propto
\frac{
\ln{
(R_{\rm max}/q_{\rm min})
}
}
{\ln{[(M_{\rm SMBH}/M_{\rm BH})(q_{\rm min}/R_{\rm max})^{1/4}]}},
\end{equation}
where $R_{\rm max}$ is the maximum integration radius, taken to correspond to the radius of the BH cluster. 
The factor in equation (\ref{loss cone log factor}) is reduced by less than 40\% 
if we consider the tidal separation of MSP-BH binaries with $a_{\rm b}=0.1$ AU instead of direct BH swallowing, for $R_{\rm max}=0.5$ pc. 
The physical reason for the weak dependence on $q_{\rm min}$ is that the process occurs in the ``empty'' loss cone regime, in which an orbiting mass is only deflected by a small amount during an orbital period. This allows it to effectively sweep over all possible eccentricities, until the narrow range corresponding to $q_{\rm min}$ is entered.

MSP-BH binaries are therefore not tidally separated in the loss cone for nearly as long as it takes to deplete stellar BH cluster. Since the time scale exceeds the age of the Universe, we neglect this process.\\ \\
\emph{Radial wandering:} Our discussion has so far implicitly assumed that binaries evolve at fixed radii from the central SMBH. 
In reality, most binaries have eccentric orbits whose orbital elements change with time as they undergo encounters with other stars, which causes them to wander in radius. 
In our picture of a steady-state stellar background, this does not affect the average MSP-BH creation rate.

The destruction time scale for any given system must however be weighted by the amount of time that the system spends at different radii. 
In the limit of a short wandering time scale, the cusp is well mixed, with a given MSP-BH spending a time in a shell of thickness $\Delta r$ proportional to the average number of systems in this shell, $4 \pi r^{2} n_{\rm MSP-BH}(r) \Delta r$. 
This limit is reasonable, because the orbital time is negligible, and the two-body relaxation time scale for 10 M$_{\odot}$ BHs is $\lesssim10^{9}$ yr. 
We therefore expect the MSP-BHs to generally have time to explore a range of semi-major axes and eccentricities in their orbits around the SMBH. 

Given an ionization/exchange survival time for a MSP-BH formed at radius $r$, parameterized as $t_{\rm surv}(r)\equiv t_{\rm surv,0}(r/r_{0})^{\delta}$, we ask what is the average survival time $\langle t_{\rm surv} \rangle$ accounting for radial wandering. 
Noting that the proper average is over the destruction rate, rather than the destruction time scale directly, we have:
\begin{align}
\langle t_{\rm surv} \rangle^{-1} = & \frac{1}{\int_{r_{\rm wand}}^{R_{\rm max}}dr r^{2} n_{\rm MSP-BH}(r)}  \\ \notag
& \times \int_{r_{\rm wand}}^{R_{\rm max}} dr r^{2} n_{\rm MSP-BH}(r) t_{\rm surv}^{-1}(r),
\end{align}
where $r_{\rm wand}$ is the minimum radius from the SMBH to which the MSP-BHs have time to wander before being destroyed. Assuming $n_{\rm MSP-BH}(r) \propto r^{-\alpha_{\rm BH}}$ (since $M_{\rm NS} \ll M_{\rm BH}$),
\begin{align}
\langle t_{\rm surv} \rangle^{-1} = & \frac{1}{\int_{r_{\rm wand}}^{R_{\rm max}}dr r^{2} r^{-\alpha_{\rm BH}}}  \\ \notag
& \times \int_{r_{\rm wand}}^{R_{\rm max}} dr r^{2} r^{-\alpha_{\rm BH}} t_{\rm surv,0}^{-1} (r/r_{0})^{-\delta}.
\end{align}
For $\delta = \alpha_{\rm BH} - 1/2$, the scaling appropriate for the dominant binary evaporation process in a Keplerian potential, and for our fiducial choice of $\alpha_{\rm BH}=1.75$, this yields
\begin{equation}
\langle t_{\rm surv} \rangle^{-1} \approx 
1.25
\left(
\frac{r_{0}}{R_{\rm max}}
\right)^{1.25}
\ln{
\left(
\frac{R_{\rm max}}{r_{\rm wand}}
\right)} t_{\rm surv,0}^{-1}
\end{equation}
for $r_{\rm wand} \ll R_{\rm max}$
To evaluate the $\ln{(R_{\rm max}/r_{\rm wand})}$ factor, we again exploit the loss cone scaling in equation (\ref{loss cone log factor}): it corresponds to the solution of the equation $t_{\rm lc} = \langle t_{\rm surv} \rangle$ with $q_{\rm min}=r_{\rm wand}$, i.e. the minimum radius that can be reached in the expected lifetime of the binary. 
Approximating the denominator of equation (\ref{loss cone log factor}) as constant yields:
\begin{align}
\ln{
\left(
\frac{R_{\rm max}}{r_{\rm wand}}
\right)}
\approx 
\left[ 
\frac{t_{\rm surv}(R_{\rm max})}{2.3{\rm~Gyr}} 
\right]^{1/2}
,
\end{align}
from which we obtain:
\begin{align}
\label{average survival time}
\langle t_{\rm surv} \rangle \approx [1.5~{\rm Gyr}\times t_{\rm surv}(R_{\rm max})]^{1/2}.
\end{align}\\
\emph{Pulsar shut off:} Although it does not destroy the NS-BH binary, if the neutron star stops emitting in the energy band of interest (e.g., in the radio), it is no longer an attractive observational target. 
The spindown time scale can be estimated as $t_{\rm spin}=P/2\dot{P}$, where $P$ is the spin period of the pulsar and $\dot{P}$ is its derivative. 
For MSPs, this time scale is typically a few Gyr. 
Since they are likely to continue emitting even after this time has elapsed, we assume that MSPs effectively live forever. 
Ordinary pulsars, on the other hand, have radio lifetimes $\lesssim100$ Myr \citep[e.g.,][]{2006ApJ...643..332F} which are too short for a significant fraction to survive long enough to undergo an exchange interaction and be observed in a BH binary created as considered here.\\ \\
\emph{Age of the Galaxy:} Regardless of the expected lifetime of the MSP-BH binary, the creation process cannot have operated for longer than the age of the Galaxy, $t_{\rm Gal}$. We therefore always cap the survival time at this value, fiducially adopting $t_{\rm Gal}=10$ Gyr.\\ \\

\subsection{MSP-WD Progenitors in the GC}
\label{MSP WD progenitors}
The creation rate and survival time considerations above require an estimate of the starting number of MSP-WDs in order to yield a prediction for the number of MSP-BHs potentially observable today. 
As this is a complex population synthesis problem in itself, we do not attempt a detailed calculation, but instead base our results on simple physical scalings. 
The starting point for our estimate are the observed MSPs in dense globular clusters, such as Terzan 5 (Ter5), whose core mass density is similar to the density at the characteristic radius of the central BH cluster. Ter5 already has over 30 pulsars detected in the radio, most of them MSPs. 
Accounting for selection effects, including the fact that the cluster has only been probed down to a finite radio flux, it could easily host 100 or more MSPs. 
Additionally, many of the globular cluster MSPs are found in binaries which frequently eclipse owing to ablation of the companion \citep[e.g.,][]{1991ApJ...379L..69T}, reducing the sensitivity of periodicity searches. 
The same MSPs would however no longer eclipse after exchanging their companion for a stellar BH, and would therefore become easier to detect.

On theoretical grounds \citep[e.g.,][]{1987IAUS..125..187V}, the number of dynamically-formed binaries should scale with the collision number
\begin{equation}
\label{collision rate}
\Gamma_{\rm c} \propto \rho_{0}^{2} r_{\rm c}^{3} \sigma_{0}^{-1},
\end{equation}
where $\rho_{0}$ is the cluster density, $r_{\rm c}$ is its characteristic radius, and $\sigma_{0}$ is its 1D velocity dispersion. 
This expectation is well supported by observations showing that the collision numbers of globular clusters correlate well with the numbers of close X-ray binaries and MSPs they host \citep[][]{2003ApJ...591L.131P, 2006ApJ...646L.143P, 2010A&A...524A..75A}.  
In extrapolating equation (\ref{collision rate}) from globular clusters to the GC, other factors must however be taken into account:\\ \\
\emph{The number of NSs per unit mass:} The high kick velocities $v_{\rm kick} \sim400$ km s$^{-1}$ of pulsars \citep[e.g.,][]{2006ApJ...643..332F} relative to the escape velocities $v_{e}\sim50$ km s$^{-1}$ of massive globular clusters imply that only a small fraction of the NSs formed in globulars should be retained. 
Assuming that the birth kick velocity is Maxwellian in form, with $\langle v_{\rm kick }\rangle=380$ km s$^{-1}$ \citep[][]{2006ApJ...643..332F}, only 0.2\% of the pulsars have $v_{\rm kick}<v_{e}=50$ km s$^{-1}$. 
In practice, the retained fraction must be $\sim10$\% in order to explain the number of NSs observed in rich clusters, with the enhanced retention fraction most likely owing to the frequent occurrence of massive binary companions around the NS progenitors \citep[e.g.,][]{1996MNRAS.280..498D, 1998MNRAS.301...15D, 2002ApJ...573..283P}. 
Within the sphere of influence of the SMBH at the GC, the escape velocity is much larger,
\begin{equation}
v_{e}(r) \approx 262{\rm~km~s^{-1}} 
\left( \frac{M}{4\times10^{6}~{\rm M_{\odot}}} \right)^{1/2}
\left( \frac{r}{0.5~{\rm pc}} \right)^{-1/2},
\end{equation}
so that a fraction of order unity of the NSs born there could be retained, i.e. up to $\sim10\times$ as many per unit mass as in massive globulars. 

Recently, observations have furthermore suggested that the initial mass function (IMF) of stars at the GC may be top-heavy \citep[][]{2010ApJ...708..834B}. 
If that is the case, a larger fraction of the main sequence stars should produce NSs than in globulars. 
Let us assume, for illustration, that practically all stars with mass $>8$ M$_{\odot}$ produce a NS upon death, with a small fraction yielding a BH instead. 
Suppose further that the IMF covers stellar masses from $M_{\rm min}=0.1$ M$_{\odot}$ to $M_{\rm max}=100$ M$_{\odot}$. 
If the IMF has an unbroken Salpeter slope, $dN/dm\propto m^{-2.35}$, each $10^{6}$ M$_{\odot}$ of star formation yields 7,400 NSs. 
If, on the other hand, the IMF is top-heavy with $dN/dm\propto m^{-0.45}$ throughout (as measured by Bartko et al. 2010 for massive stars\nocite{2010ApJ...708..834B}), the same total mass of star formation produces 21,200 NSs, nearly $3\times$ as many as the fiducial Salpeter case. 
In the intermediate scenario in which the IMF has a Salpeter slope below 8 M$_{\odot}$ and -0.45 above, 16,300 NSs are produced, an enhancement of a factor of 2.2 over the pure Salpeter IMF. 
Combining a possibly more efficient NS production rate and a higher retention fraction, the GC could therefore be richer in NSs than globular clusters, per unit mass, by as much as a factor of $\sim20-30$.\\ \\
\emph{Binary fraction:} The dynamical interactions considered in this work can only operate in the GC if binaries are present. 
There are currently almost no direct constraints on the binary fraction in the GC, precluding estimates at better than the order-of-magnitude level. 
Observations of hypervelocity stars (HVSs) (which in the classical model arise from the dynamical interaction of binaries with the central SMBH; Hills 1988\nocite{1988Natur.331..687Hls}), of the S-stars around Sgr A$^{\star}$ (which may be the counterparts of HVSs captured by the SMBH; Gould \& Quillen 2003\nocite{2003ApJ...592..935G}, Ginsburg \& Loeb 2006\nocite{2006MNRAS.368..221G}), and the detection of several X-ray binaries with projected separation $<$1 pc from Sgr A$^{\star}$ (including one at $<0.1$ pc; Muno et al. 2005a,b, Bower et al. 2005\nocite{2005ApJ...622L.113M, 2005ApJ...633..228M, 2005ApJ...633..218B}) however make a strong case that binaries do exist in non-negligible numbers in the GC. 
Assuming that $\sim10\%$ of stars are members of a binary with semi-major axis $a_{\rm b}\lesssim0.3$ AU, as in the Galactic disk \citep[e.g.,][]{1991A&A...248..485D} and similar to what is inferred for globular clusters \citep[e.g.,][]{1992PASP..104..981H, 1997ApJ...474..701R}, \cite{2003ApJ...599.1129Y} predicted a HVS production rate roughly consistent with what is observed \citep[][]{2007ApJ...671.1708B}. 
It is therefore reasonable to assume that the GC is not severely depleted in binaries. 
\\ \\
Let $f_{\rm NS}$ be the factor by which the GC retains more NSs per unit stellar mass than massive globulars, and let $f_{\rm bin}$ be the ratio of the binary fractions at the GC and in globulars. 
We account, roughly, for the fact that binaries may be ionized down to smaller semi-major axes in the GC relative to globular clusters (as a result of the higher velocity dispersion) by adopting $f_{\rm bin}=0.5$. 
For a binary semi-major axis distribution that is flat in the logarithm, the effect of truncating the upper end should in fact be modest on the total number of binaries. 
From the above considerations, we have:
\begin{equation}
\frac{N_{\rm MSP}^{\rm GC}}
{N_{\rm MSP}^{\rm gl}} \sim 10
\left( \frac{\rho_{0}^{\rm GC}}{\rho_{0}^{\rm gl}} \right)^{2}
\left( \frac{r_{c}^{\rm GC}}{r_{c}^{\rm gl}} \right)^{3}
\left( \frac{\sigma_{0}^{\rm GC}}{\sigma_{0}^{\rm gl}} \right)^{-1}
\left( \frac{f_{\rm NS}}{20} \right)
\left( \frac{f_{\rm bin}}{0.5} \right),
\end{equation}
where `gl' stands for `globular cluster' and `GC' stands for `Galactic center' as before. 
Here, $N_{\rm MSP}^{\rm GC}$ and $N_{\rm MSP}^{\rm gl}$ are the numbers of MSPs, assuming the same total stellar mass. 

For this comparison, we adopt the globular cluster parameters of Ter 5, $\rho_{0}^{\rm gl}=10^{6}$ M$_{\odot}$ pc$^{-3}$, $r_{\rm c}^{\rm gl}=0.1$ pc, $\sigma_{0}^{\rm gl}=12$ km s$^{-1}$ \citep[][]{1985IAUS..113..541W, 2002ApJ...571..818C}. 
For the Galactic center, we assume a characteristic radius $r_{\rm c}^{\rm GC}=0.5$ pc for the cluster of stellar BHs, and evaluate the stellar density and velocity dispersion at this radius.
From equations (\ref{stellar density profile}) and (\ref{velocity dispersion}), and for the GC model described in \S \ref{stellar distribution}, we find $\sigma_{0}^{\rm GC}=139$ km s$^{-1}$ and $\rho_{0}^{\rm GC}=3.3\times10^{5}$ M$_{\odot}$ pc$^{-3}$. 
This yields:
\begin{equation}
\frac{N_{\rm MSP}^{\rm GC}}
{N_{\rm MSP}^{\rm gl}} \sim 12
\left( \frac{f_{\rm NS}}{20} \right)
\left( \frac{f_{\rm bin}}{0.5} \right).
\end{equation}
Since Ter 5 and the central parsec both contain a stellar mass $\approx10^{6}$ M$_{\odot}$, the above estimate implies that the central parsec could contain as many as 12$\times$ more MSPs than Ter 5, i.e., up to 1,200 if Ter 5 hosts 100 of them. 

There are important uncertainties in extrapolating from globular clusters to the Galactic center. 
We therefore caution the reader against over-interpreting this crude calculation. 
Nevertheless, assuming that the MSP formation channels operating in globulars operate in the GC as well \citep[as supported by the observed overabundance of X-ray binaries in the central parsec; e.g.,][]{2005ApJ...622L.113M}, several hundreds to more than a thousand MSPs could populate the nuclear BH cluster today. 
We will adopt a nominal value $N_{\rm MSP}=500$ in the estimates that follow, approximately in the middle of the expected range. 
This number consistent with the upper limit of 90 ordinary pulsars in the central parsec reported by \cite{2010ApJ...715..939M}, as their observations were insensitive to MSPs due to scattering pulse broadening at their observing frequency (15 GHz). 
Furthermore, globular clusters typically contain many more MSPs than ordinary pulsars. 
As recently shown by \cite{2010arXiv1011.4275A}, the $\gamma-$ray flux and spectrum measured by the Fermi Gamma-ray Space Telescope\footnote{http://fermi.gsfc.nasa.gov} (\emph{Fermi}) toward the GC (over a larger, 175 pc radius) is also consistent with a enhancement of MSPs in this region. 

For our fiducial parameters, we finally find
\begin{equation}
\label{expected N MSP BH}
N_{\rm MSP-BH}^{\rm surv} \approx 
6 \left( \frac{N_{\rm MSP}}{500}
\right).
\end{equation}

\begin{table*}
\centering
\caption{Dependence of the Number of MSP-BHs Surviving in the Galactic Center Today on Model Parameters\label{params table}}
\begin{tabular}{|ccccccccccccc|}
\hline\hline
Parameter Varied & $R_{\rm min}$ & $R_{\rm max}$ & $\alpha_{\rm BH}$ & $\alpha_{\rm MSP}$ &  $M_{\rm WD}$ & $M_{\rm BH}$ & $N_{\rm BH}$ & $a_{\rm b,min}$  & $a_{\rm b,max}$ & $t_{\rm Gal}$ & $N^{\rm surv}_{\rm MSP-BH} (500/N_{\rm MSP})$   \\
            & pc                      & pc                         &                                  &                                        & $M_{\odot}$      & $M_{\odot}$    &                           & AU                         & AU                         & Gyr                       &\\
\hline
Fiducial                       &  0.1                   &  0.5                       & 1.75                           & 1.3                   & 1                        & 10                      & 25,000                 & 0.01                       & 0.3                        &  10                & 6 \\ 
$R_{\rm min}$          & \bf{0.01}           &  0.5                       & 1.75                           & 1.3                   & 1                        & 10                      & 25,000                 & 0.01                        & 0.3                        &  10                & 7 \\ 
$R_{\rm min}$          & \bf{0.2}             &  0.5                       & 1.75                           & 1.3                   & 1                         & 10                     & 25,000                 & 0.01                        & 0.3                        &   10               & 6 \\ 
$R_{\rm max}$         & 0.1                    & \bf{1}                    & 1.75                           & 1.3                   &  1                        & 10                     & 25,000                 & 0.01                        & 0.3                        &   10               & 6 \\ 
$\alpha_{\rm BH}$    & 0.1                    &  0.5                       & \bf{1.5}                      &  1.3                   &   1                       & 10                     & 25,000                & 0.01                         & 0.3                        &   10               & 12 \\ 
$\alpha_{\rm MSP}$ & 0.1                    &  0.5                       & 1.75                           &  \bf{1.5}          &  1                         & 10                     & 25,000                & 0.01                        & 0.3                        &  10                & 7 \\ 
$M_{\rm WD}$           &  0.1                   &  0.5                       & 1.75                           & 1.3                   & \bf{0.5}             & 10                      & 25,000                 & 0.01                        & 0.3                        &  10                & 4 \\ 
$M_{\rm WD}$          &  0.1                   &  0.5                       & 1.75                           & 1.3                   & \bf{0.2}             & 10                      & 25,000                 & 0.01                        & 0.3                        &  10                & 3 \\ 
$M_{\rm BH}$           &  0.1                   &  0.5                       & 1.75                           & 1.3                   & 1                        & \bf{7}                 & 25,000                 & 0.01                        & 0.3                        &  10                & 5 \\ 
$M_{\rm BH}$           & 0.1                    &  0.5                       & 1.75                           & 1.3                   & 1                        & \bf{13}               & 25,000                 & 0.01                        & 0.3                        &  10                & 8 \\ 
$N_{\rm BH}$           & 0.1                    &  0.5                       & 1.75                           & 1.3                   &  1                        & 10                     & \bf{15,000}         & 0.01                         & 0.3                        &  10                & 5 \\ 
$N_{\rm BH}$            & 0.1                    &  0.5                       & 1.75                           & 1.3                   &  1                        & 10                     & \bf{20,000}         & 0.01                         & 0.3                        & 10                  & 6 \\ 
$a_{\rm b,min}$        & 0.1                    &  0.5                       & 1.75                           &  1.3                 &  1                          & 10                     & 25,000               & \bf{0.05}                 & 0.3                         &  10                & 6 \\ 
$a_{\rm b,min}$        & 0.1                    &  0.5                       & 1.75                           &  1.3                 &   1                         & 10                     & 25,000               & \bf{0.1}                   & 0.3                        & 10                 & 7 \\ 
$a_{\rm b,max}$       & 0.1                   &  0.5                       & 1.75                           &  1.3                 &   1                          & 10                    & 25,000               & 0.01                        & \bf{1}                    &  10                & 5 \\ 
$a_{\rm b,max}$       & 0.1                   &  0.5                       & 1.75                           &  1.3                 &   1                          & 10                    & 25,000               & 0.01                        & \bf{3}                     &  10               & 5 \\ 
$t_{\rm Gal}$             &  0.1                   &  0.5                       & 1.75                           & 1.3                   & 1                        & 10                      & 25,000                 & 0.01                       & 0.3                        &  \bf{5}          & 6 \\ 
$t_{\rm Gal}$             &  0.1                   &  0.5                       & 1.75                           & 1.3                   & 1                        & 10                      & 25,000                 & 0.01                       & 0.3                        &  \bf{2}          & 4 \\ 
\hline
\tablecomments{All symbols are defined in the text. The fiducial parameter values are given in the top row and the model parameters are varied one at a time relative to the fiducial set, with the variation indicated in bold. All other parameters, such as the mass of the central massive black hole and the total mass distribution in the nuclear star cluster, are held fixed.}
\end{tabular}
\end{table*}

\subsection{Sensitivity to Model Parameters}
\label{sensitivity to model parameters}
The calculations presented above assumed fiducial parameters for definiteness. 
In order to estimate the sensitivity of our results to the model parameters, we have repeated the calculations varying key ones, one at a time. 
Specifically, we have explored the variations listed in Table \ref{params table}, in which we have varied each parameter by amounts representative of the expected range (for the case $\alpha_{\rm BH}\neq1.75$, the survival time is evaluated using a generalization of eq. (\ref{average survival time})). 
In addition to actual physical parameters, we have varied the Galaxy ``age'' $t_{\rm Gal}$ as a test of the steady-state approximation. 
In all cases, the predicted number of surviving MSP-BHs is within a factor of 2 of the fiducial result in equation (\ref{expected N MSP BH}). 
At the order-of-magnitude level of the present work, our estimates are therefore robust to the details of the assumptions. 

The white dwarf mass, $M_{\rm WD}$, warrants further discussion, as it also affects the predicted properties of the MSP-BH systems. 
In fact, the dimensionless ratio $M_{\rm WD}/M_{\rm BH}$ was shown in \S \ref{survival time} to determine both the factor by which the semi-major axis is multiplied during the exchange interaction and the eccentricity of the resulting MSP-BH. 
While the black hole mass also appears in this ratio, observations suggest that they have a narrower mass distribution \citep[e.g.,][]{2010arXiv1006.2834O}. 
The fiducial choice of $M_{\rm WD}=1$ M$_{\odot}$, on the other end, is at the upper end of the observed white dwarf mass distribution, with values as low as $M_{\rm WD}\sim0.1$ M$_{\odot}$ also occurring (even lower mass WD companions have been observed [e.g. Stairs 2004\nocite{2004Sci...304..547S}], but they are rare). 
Somewhat higher eccentricities than for the fiducial case are therefore possible, e.g. $e'\sim0.98$ for $M_{\rm WD}=0.1$ M$_{\odot}$, although the scatter in eccentricities resulting from exchange interactions is sufficiently large that this value would actually be perfectly compatible with $M_{\rm WD}=1$ M$_{\odot}$ as well \citep[e.g.,][]{1993ApJ...415..631S}. 
In principle, MSP-BH semi-major axes larger than the $\sim3$ AU expected from a $M_{\rm BH}/M_{\rm WD}=10$ exchange with a binary with original semi-major axis $a_{\rm b,max}=0.3$ AU are also possible for $M_{\rm BH}/M_{\rm WD}\gg10$, but these should be quickly evaporated (Fig. \ref{3body time scales}).

\section{DISCUSSION}
\label{discussion}

\subsection{Observational Prospects}
\label{observational prospects}
Our results suggest that our Galactic center should harbor several MSP-BH binaries, providing a strong motivations for focused searches in this direction. 
As no pulsar has so far been detected in the central parsec, it is important to consider whether finding MSP-BHs there is feasible.

The time scale by which a radio pulse is broadened owing to interstellar scattering is 
\begin{equation}
\tau_{\rm scat} = 116~{\rm ms}
\left( \frac{D_{\rm scat}}{100~{\rm pc}} \right)^{-1}
\left(
\frac{\nu}{10~{\rm GHz}}
\right)^{-4},
\end{equation}
where $D_{\rm scat}$ is the distance of the effective scattering screen from the GC.
Combining all known tracers of ionized gas, \cite{1998ApJ...505..715L} found $D_{\rm scat}=133^{+200}_{-80}$ pc; we assume $D_{\rm scat}=100$ pc. 
Sensitivity to a pulsed signal of period $P$ requires $\tau_{\rm scat}\lesssim P$. 
For a period $P=5$ ms at the GC, the minimum observing frequency is therefore $22$ GHz, while $33$ GHz is sufficient to detect a 1 ms source. 
Let us consider a hypothetical 30 GHz search, which would suffice to detect most MSPs if sufficiently deep. 
The minimum flux density for a pulsar detection is 
\begin{equation}
\label{radiometer equation}
S_{\rm min}=\delta_{\rm beam} \frac{\beta_{\rm sys} \sigma (T_{\rm rec}+T_{\rm sky})}{G\sqrt{N_{p}\Delta \nu t_{\rm int}}}\sqrt{\frac{W_{e}}{P-W_{e}}},
\end{equation}
where $\delta_{\rm beam}$ is a factor accounting for the reduction in sensitivity to pulsars located away from the center of the telescope beam, $T_{\rm rec}$ is the receiver temperature on a cold sky, $T_{\rm sky}$ is the sky background temperature, $G$ is the antenna gain, $N_{p}$ is the number of polarizations summed, $\Delta \nu$ is the receiver bandwidth, $t_{\rm int}$ is the integration time, $P$ is the pulse period, $W_{e}$ is the effective pulse width, $\sigma$ is the signal-to-noise (S/N) detection threshold, and $\beta$ is a constant accounting for various system losses \citep[e.g.,][]{dss+84}. 
\cite{2010ApJ...715..939M} calculated the S/N expected as a function of observing frequency for a 10-hour observations with the GBT with a bandwidth $\Delta \nu = 800$ MHz, for different pulsar periods and spectral properties. 
We can use equation (\ref{radiometer equation}) to scale their results and explore the potential capabilities of future searches. 

\cite{2010ApJ...715..939M} predict a $S/N=1$ for a 10-hour, 30 GHz GBT integration with $\Delta \nu=800$ MHz, for a 5-ms pulsar with a typical spectral index $\alpha=-1.7$ ($S_{\nu}\propto \nu^{\alpha}$) and flux density of 1 mJy at 1 GHz (corresponding to $L_{1000}=S_{1000}d^{2}=72.25$ mJy kpc$^{2}$ at the GC, $d=8.5$ kpc), and $S/N=10$ if the spectral index is instead -1. 
For reference, the observed spectral index distribution is well modeled by a normal distribution with mean $-1.7$ and dispersion $0.35$ \citep[e.g.,][]{2009A&A...505..919S}. 
To get a sense of the limits of a dedicated GBT campaign with upgraded instrumentation, we consider a 100-hr integration with a $\Delta \nu=8$ GHz bandwidth instead\footnote{A spectrometer capable of such high-frequency, large bandwidth observations is already being developed for the GBT (http://www.gb.nrao.edu/gbsapp/).}, in which case the same S/Ns are achieved for a pulsar with flux density lower by a factor of 10. 
A futuristic Square Kilometer Array\footnote{http://www.skatelescope.org} (SKA) telescope would have an effective area larger by a factor of $\sim100$, and hence would be sensitive to pulsars $\sim100\times$ fainter still. 
Thus, an upgraded GBT campaign could in principle detect a $L_{1000}\approx7$ mJy kpc$^{2}$, $\alpha=-1$, 5-ms MSP at the GC with $S/N=10$, while the SKA could in principle probe as faint as $L_{1000}\approx0.7$ mJy kpc$^{2}$ under the same conditions. 
For a more typical spectral index $\alpha=1.7$, the same S/N is reached for pulsars more luminous at 1 GHz by a factor of 10. 
For comparison, the $least~luminous$ MSPs currently detected in deep globular cluster observations have $L_{1000}\sim1$ mJy kpc$^{2}$ and the luminosity function extends beyond at least $L_{1000}\sim100$ mJy kpc$^{2}$ \citep[e.g.,][]{2007ApJ...670..363H}. 
A dedicated GBT search should therefore be capable of detecting MSPs at the GC, while the SKA could potentially probe a substantial fraction of the population. 

\emph{Fermi} has demonstrated that MSPs are $\gamma-$ray sources \citep[e.g.,][]{2009Sci...325..848A, 2010JCAP...01..005F}. 
The $\gamma-$ray energy band has the advantage that, unlike the radio, it is not affected by interstellar scattering. 
However, even \emph{Fermi} can only detect the very brightest, tip-of-the-iceberg $\gamma-$ray pulsars at the GC, so it is unlikely that a significant number of individual $\gamma-$ray MSPs will detected in the central parsec for the foreseeable future.

\subsection{Signatures of the Proposed Formation Channel}
\label{signatures}
Our calculations make definite predictions for the properties of the PSR-BHs formed via the proposed channel:
\begin{enumerate}
\item The system would be found within $\sim1$ pc of Sgr A$^{\star}$, where the cluster of stellar BHs is assumed to be located. 
\item The pulsar would be recycled, with a period from $\sim1$ to a few tens of milliseconds, in order to have an original WD companion and to have a sufficiently long spin down time to remain emitting in the radio to the present day. 
Accordingly, the pulsar should have a low magnetic field, $B\lesssim10^{10}$ G.
\item The MSP-BH binary would be relatively wide, as the semi-major axis is multiplied by a factor $\sim M_{\rm BH}/M_{\rm WD}\gtrsim10$ during the exchange interaction. For an original MSP-WD semi-major axis distribution ranging from $\sim0.01$ AU to $\sim0.3$ AU, the MSP-BHs should have semi-major axes ranging from $\sim0.1$ to $\gtrsim3$ AU. \item The MSP-WD would be highly eccentric, $1-e'\sim(M_{\rm WD}/M_{\rm BH})(M_{\rm BH}/M_{\rm NS})^{1/3}$ (\S \ref{survival time}). For our fiducial masses, $e'\sim0.8$, but the eccentricity distribution will be broad.
\end{enumerate}

These predictions are based on the properties of the MSP-BHs immediately after their formation in exchange interactions. 
In principle, their internal properties could evolve by the time they are detected. 
Simple estimates however suggest that they should not change qualitatively. 
As discussed in \S \ref{survival time}, the spin down time scales of recycled pulsars are typically of order of a few Gyr. 
In the soft regime, the binary evaporation time $t_{\rm evap}\propto a_{\rm b}'^{-1}$ (eq. (\ref{evaporation time scale})), and so the binaries spend most of their lifetime with a semi-major axis of the order of its value at formation. 
In the hard regime, the binary semi-major axis is progressively reduced by hardening, but the time scale for this process is comparable to the exchange interaction time scale \citep[e.g.,][]{1993ApJS...85..347H}, which for MSP-BHs at the GC is $\gtrsim10$ Gyr (Fig. \ref{3body time scales}). 
These simple considerations are in good agreement with the Fokker-Planck modeling of binaries at the GC by \cite{2009ApJ...700.1933H}, who found little evolution of the internal binary properties. Thus, ``Heggie's law'' (according to which soft binaries become softer and hard binaries become harder) holds but the time scales are too long for it to have a large impact.

We use the cross section derived by \cite{1996MNRAS.282.1064H} to estimate the time it takes for encounters with other stars to change the eccentricity by an amount $>\delta e_{0}$, starting with a value $e$. 
Averaging over a Maxwellian velocity distribution,
\begin{align}
t_{e}(>\delta e_{0})
\approx 
0.29
\frac{\sigma_{3}}{G n_{3} M_{123} a_{b}'}
& \left( \frac{M_{12}M_{123}}{m_{3}^{2}} \right)^{1/3} \notag \\
& \times \frac{\delta e_{0}^{2/3}}{e^{2/3}(1 - e^{2})^{1/3}},
\end{align}
where $M_{12}\equiv m_{1}+m_{2}$ and $M_{123}\equiv m_{1}+m_{2}+m_{3}$. Comparing with the ionization time scale given by equation (\ref{ionization time scale}), we find
\begin{equation}
\frac{t_{e}(>\delta e_{0})}{t_{\rm ion}} 
\approx 9.69
\frac{m_{3}^{4/3}}{M_{123}^{2/3}M_{12}^{2/3}} 
\frac{\delta e_{0}^{2/3}}{e^{2/3}(1 - e^{2})^{1/3}}.
\end{equation}
For our fiducial MSP-BH mass choices, $e=0.8$ and $\delta e_{0}=0.1$, $t_{e}(>\delta e_{0})/t_{\rm ion}\approx1.9$, indicating that the eccentricity changes should also be modest over the lifetimes of the binaries. 
It is also noteworthy that even if eccentricity perturbations occurred on a shorter time scale, the systems would be asymptotically driven to a thermal distribution, $f_{\rm th}(e)=2e$ with mean $\langle e \rangle_{\rm th} = 0.67$, so that the eccentricities would remain large on average. 

\subsection{Applications}
\label{applications}
The discovery of a MSP-BH system at the GC would have profound implications. 
At this time, it would be the first known PSR-BH binary and would therefore critically inform our theories of stellar evolution, especially in dense stellar environments. 
If it has the properties outlined above, it would convincingly originate in an exchange scenario similar to the one we have explored.
It would therefore provide strong evidence for the existence of the predicted cluster of stellar BHs around Sgr A$^{\star}$, for which there is currently no observational confirmation. 
This would in turn have important implications for a number of other phenomena, including HVSs \citep[e.g.,][]{2008MNRAS.383...86O}, gravitational wave sources \citep[particularly for the extreme mass ratio events to be detected by \emph{LISA}\footnote{http://lisa.nasa.gov} but also for \emph{LIGO}\footnote{http://www.ligo.caltech.edu}; e.g.,][]{2007PhRvD..75d2003B, 2009MNRAS.395.2127O, 2010arXiv1010.5781A}, microlensing near Sgr A$^{\star}$ \citep[e.g.,][]{2001ApJ...563..793C, 2001ApJ...551..223A}, and the orbital capture of stars by Sgr A$^{\star}$ \citep[e.g.,][]{2004ApJ...606L..21A}.

It would also for the first time provide an accurate clock orbiting a BH and therefore offer a unique probe of the spacetime in a BH potential \citep[e.g.,][]{2003LRR.....6....5S, 2007mru..confE..20K}. 
While the MSP-BHs formed in this scenario would not be as tightly bound as the relativistic double NS systems from which the best constraints on gravity theories are currently derived \cite[e.g.,][]{2006Sci...314...97K, 2010ApJ...722.1030W}, certain relativistic effects including the periastron shift and the Shapiro time delay can be measured even in relatively loose systems \citep[e.g.,][]{1988ApJ...332..770T, 1991ApJ...371..739R, 1991ApJ...379L..17N}. 
Measuring these effects would provide much more accurate measurements of the masses of stellar BHs than currently possible in X-ray binaries, in which assumptions regarding the inclination of the binary must be made. 
Another interesting possibility for nearly edge-on systems would be to measure the gravitational lensing effects of the BH using pulse delays, from which the orientation and spin of the BH could be inferred \citep[e.g.,][]{1991ApJ...379L..17N, 2006ApJ...636L.109B}. 

\subsection{Caveats}
\label{caveats}
While our calculations provide straightforward predictions for where and how to find MSP-BH binaries, of their expected properties, and suggest that they should be observationally accessible, they are limited in some respects. 
As a first investigation of the proposed formation channel, we have focused on simple analytic estimates of the relevant physical processes, and modeled the GC as a steady-state background in which MSP-BHs are created and destroyed. 
In reality, the GC is a dynamical system in which complicated stellar evolution processes are ongoing. 
It would therefore be desirable, as the capabilities become available in the future, to perform more detailed dynamical simulations that include the critical binary and stellar evolution effects. 

Some of the assumptions we have made have also not yet been empirically confirmed. 
In particular, our results rely on the presence of a central cluster of $\sim25,000$ stellar BHs induced by mass segregation, but this cluster may not be present if relaxation is too inefficient or if a binary massive black hole recently destroyed it \citep[e.g.,][]{2010ApJ...718..739M}. 
While the disk morphology of the Milky Way argues against a major merger in the last $\sim10$ Gyr, intermediate mass black holes brought in by dwarf galaxies or globular clusters cannibalized by the bulge could also eject stellar BHs from the central cluster. Resonant relaxation, which we have mostly neglected, could furthermore be more efficient at depleting the stellar BH cluster or binaries near Sgr A$^{\star}$ than we have assumed. 
It will be important to consider these issues in more detail in the future, although detecting MSP-BHs in the central parsec would actually inform these open questions.

\subsection{Conclusion}
\label{conclusion}
We have shown that if dynamical processes analogous to those operating in dense globular clusters occur in the central parsec of our Milky Way, and if this region hosts a cluster of $\sim25,000$ stellar-mass BHs as predicted by mass segregation arguments, then MSP-BH binaries should be formed there in exchange interactions. 
Taking into account the much higher retention fraction of neutron stars in the Galactic center relative to globular clusters, as a result of the deeper potential well, several of these systems should survive to the present day.
Simple scaling arguments indicate that some of these MSP-BHs might be detectable with existing radio telescopes, and that a substantial fraction of the population should be accessible to the Square Kilometer Array. 
Our predictions therefore suggest a definite roadmap to the detection of the first pulsar-black hole binary, by singling out a small region of the Galaxy where many might reside. 
In light of the remarkable potential payoffs of such a discovery, it is therefore clear that focused observational searches and further theoretical studies are warranted.

\section*{Acknowledgments} 
We are grateful to Ryan O'Leary and Nick Stone for useful discussions about dynamics in the Galactic center, and to Hagai Perets for discussions regarding stellar populations in the central parsec. 
We also thank the referee, Pau Amaro-Seoane, for detailed comments. 
CAFG is supported by a fellowship from the Miller Institute for Basic Research in Science, and was further supported by a FQRNT Doctoral Fellowship and a Harvard Merit Fellowship during the course of this work. 
This work was supported in part by NSF grant AST-0907890 and NASA grants
NNX08AL43G and NNA09DB30A.

\bibliography{references} 

\end{document}